\documentclass[prl,twocolumn,amsfonts,amssymb,amsmath,showpacs,floatfix,superscriptaddress]{revtex4}
\usepackage{graphicx}

\begin{document}

\title{Unconditional two-mode squeezing of separated atomic ensembles}
\author{A.~S. Parkins}
\altaffiliation[Permanent address: ]{Department of Physics, University of Auckland, 
Private Bag 92019, Auckland, New Zealand}
\affiliation{Max-Planck Institut f\"ur Quantenoptik, Hans-Kopfermann-Strasse 1,
D-85748 Garching, Germany}
\affiliation{Institut f\"ur Theoretische Physik, Universit\"at Innsbruck, and 
Institut f\"ur Quantenoptik und Quanteninformation, \"Osterreichische Akademie der Wissenschaften, A-6020 Innsbruck, Austria}
\author{E. Solano}
\affiliation{Max-Planck Institut f\"ur Quantenoptik, Hans-Kopfermann-Strasse 1,
D-85748 Garching, Germany}
\affiliation{Secci\'on F\'isica, Departamento de Ciencias, Pontificia Universidad Cat\'olica del Per\'u, Apartado Postal 1761, Lima, Peru}
\author{J.~I. Cirac}
\affiliation{Max-Planck Institut f\"ur Quantenoptik, Hans-Kopfermann-Strasse 1,
D-85748 Garching, Germany}
\date{\today}

\begin{abstract}
We propose schemes for the unconditional preparation of a two-mode squeezed
state of effective bosonic modes realized in a pair of atomic ensembles 
interacting collectively with optical cavity and laser fields. The scheme uses Raman 
transitions between stable atomic ground states and under ideal conditions 
produces pure entangled states in the steady state. The scheme works both for
ensembles confined within a single cavity and for ensembles confined in 
separate, cascaded cavities.
\end{abstract}
\maketitle

Atomic ensembles are currently attracting considerable theoretical and 
experimental interest from the quantum optics and quantum information communities
\cite{Lukin03,Kuzmich03a,Duan00,Julsgaard01,DiLisi02,Lukin00a,%
Lukin00b,Mewes05,Phillips01,Liu01,Eisaman04,Chaneliere05,Duan01,Kuzmich03b,Matsukevich04,%
Chou05,Andre02,Sorensen02,Hammerer04,Dantan05,Geremia04,Berry02,DiLisi05}. 
Collective enhancement of their interaction with electromagnetic fields enables
efficient and controllable coupling to (few-photon) non-classical light fields without 
the need for strong single-photon single-atom coupling. Given long atomic ground-state 
coherence lifetimes, they also offer a robust
medium for long-lived, high-fidelity storage of quantum states, i.e., for quantum 
memory. Of particular interest in this context is the preparation of long-lived 
quantum entangled states of two or more {\em separate} atomic ensembles
\cite{Duan00,Julsgaard01,DiLisi02,Duan01,Chou05,Berry02,DiLisi05,Lukin00b,Dantan05}, 
with the possibility of application to quantum communication protocols such as 
quantum teleportation \cite{Dantan05}. 

To date, schemes for preparing entangled states of separate atomic ensembles have
generally been based either on projective measurements
\cite{Duan00,Julsgaard01,DiLisi02,Duan01,Chou05,Lukin00b} and possibly feedback
\cite{Berry02,DiLisi05}, or on the transfer of
quantum statistics from quantum-correlated light fields \cite{Lukin00b,Dantan05}. 
Here we propose a scheme
which requires neither of these; based on a form of quantum reservoir engineering,
it is able to produce pure entangled (two-mode squeezed)
states of separate atomic ensembles in 
{\em steady state}. Consideration of potential experimental parameters suggests
that this scheme is feasible with existing experimental capabilities and could 
produce high degrees of entanglement on timescales which are orders
of magnitude shorter than achievable coherence lifetimes in atomic ensembles
\cite{Julsgaard01,Kuhr03}.

\begin{figure}[h]
\begin{center}
\includegraphics[width=2.8in]{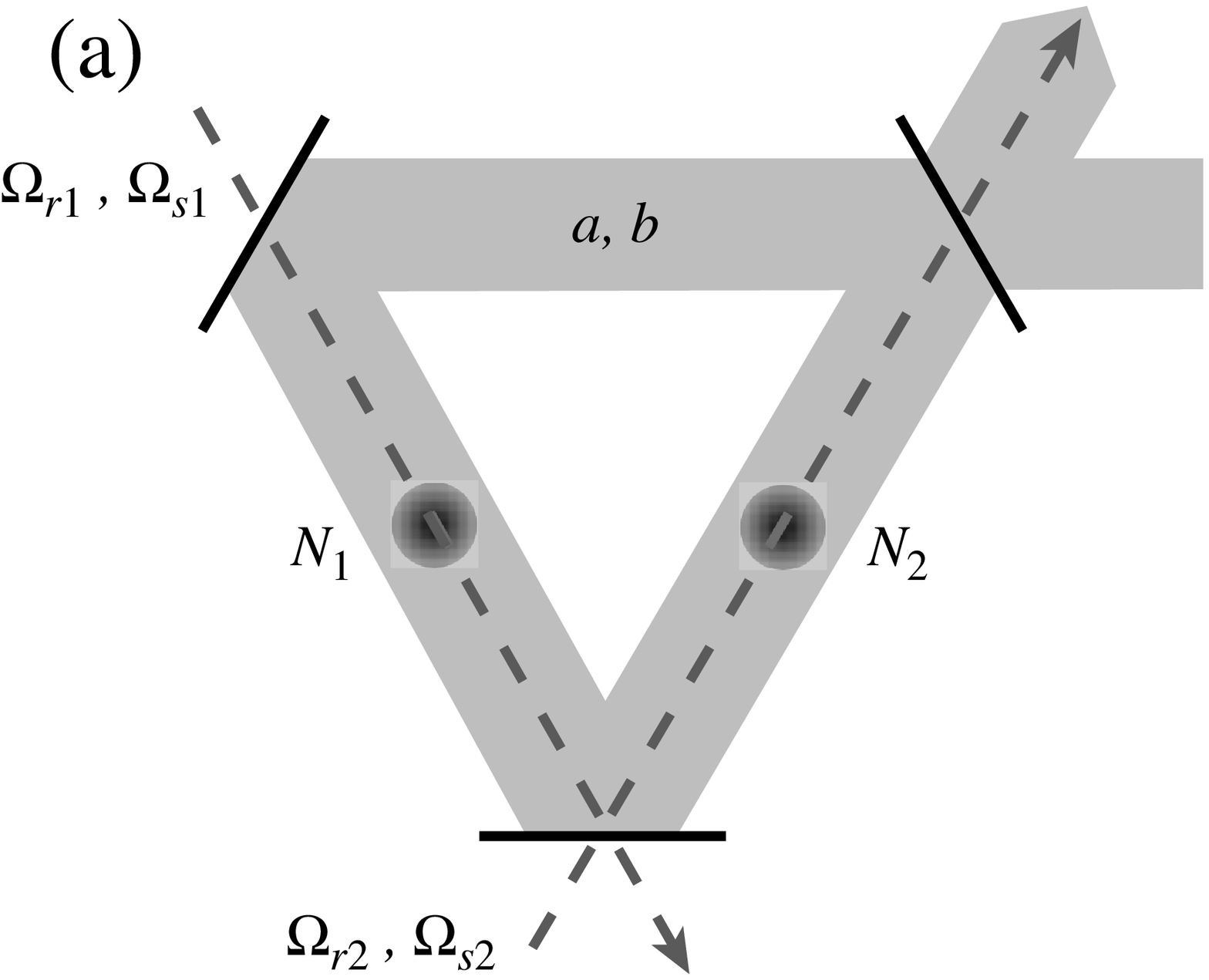}\\
\includegraphics[width=2.8in]{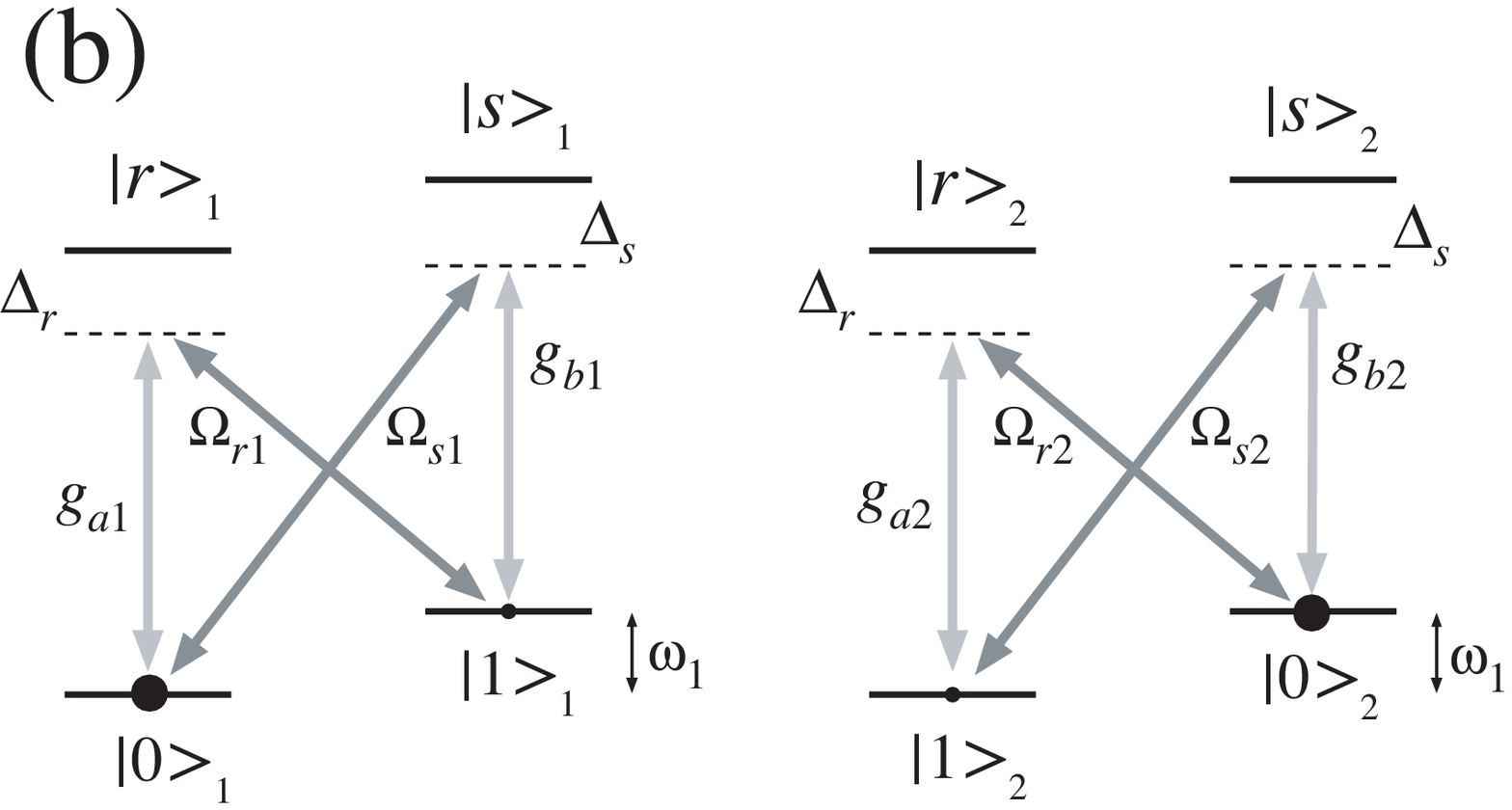}
\caption{
(a) Possible ring cavity setup. Ensembles 1 and 2 contain $N_1$ and $N_2$ atoms, 
respectively. 
(b) Atomic energy levels and excitation schemes for ensembles 1 (left) and 2 (right). 
Excited states $|r\rangle$ and $|s\rangle$ can be replaced by a 
single level, provided the two Raman channels
remain distinct.}
\label{fig:levels}
\end{center}
\end{figure}

Our proposed scheme is illustrated schematically in Fig.~1. Two orthogonal 
traveling-wave cavity modes (annihilation operators $a$ and $b$)
couple to atomic transitions with strengths $g_{ai}$ 
and $g_{bi}$, respectively, where $i=1,2$ denotes the particular atomic ensemble. 
Classical laser fields, with Rabi frequencies
$\{\Omega_{ri},\Omega_{si}\}$, combine with the cavity fields to drive two distinct Raman 
transitions between the atomic ground states $|0\rangle_i$ and $|1\rangle_i$ \cite{Guzman05}.
With a co-propagating field geometry as shown in Fig.~1, the (first-order) Doppler
effect is eliminated and, provided the light beams are broad in width compared to 
the ensembles, we can assume a uniform coupling strength 
of the atoms to each of the fields.
The two ensembles are initially prepared via separate optical pumping
in different ground states, but for convenience we relabel the ground states in 
ensemble 2 so that all atoms are initially in state $|0\rangle$ in our theoretical
treatment.

Denoting the laser frequencies by $\omega_{Ls}$ and $\omega_{Lr}$, we consider
the case where
$\omega_{Ls}-\omega_1=\omega_{Lr}+\omega_1$ \cite{tuning} and, assuming large detunings
$\Delta_r$ and $\Delta_s$ of the fields from the atomic transition frequencies,
we perform a standard adiabatic elimination of the atomic excited states and 
neglect atomic spontaneous emission. Defining collective atomic spin operators
by
\begin{eqnarray}
J_{zi} = \frac{1}{2}\sum_{j=1}^{N_i}\left( |1\rangle\langle 1|_i^j - |0\rangle
\langle 0|_i^j \right) , ~~~
J_i^- = \sum_{j=1}^{N_i} |0\rangle\langle 1|_i^j \, ,
\end{eqnarray}
the master equation for the density operator of the total system 
can then be written (after unitary transformation to an appropriate rotating frame) as
\begin{eqnarray} \label{MEab}
\dot{\rho} &=& -{\rm i}\left[ H_{\rm eff},\rho\right] + \kappa_a D[a]\rho 
+ \kappa_b D[b]\rho ,
\end{eqnarray}
where $D[{\cal O}]\rho\equiv 2{\cal O}\rho {\cal O}^\dag -{\cal O}^\dag {\cal O}\rho -\rho {\cal O}^\dag {\cal O}$,
$\{\kappa_a,\kappa_b\}$ are the cavity field decay rates, and 
\begin{widetext}
\begin{eqnarray}
H_{\rm eff} &=& \left[ \delta_a + \frac{\left| g_{a1}\right|^2}{\Delta_r}\left( 
\frac{N_1}{2} - J_{z1}\right) +  \frac{\left| g_{a2}\right|^2}{\Delta_r}
\left( \frac{N_2}{2} + J_{z2} \right) \right] a^\dag a  
+  \left[ \delta_b + \frac{\left| g_{b1}\right|^2}{\Delta_s}\left( \frac{N_1}{2} + J_{z1}
\right) +  \frac{\left| g_{b2}\right|^2}{\Delta_s} \left( 
\frac{N_2}{2} - J_{z2} \right) \right] b^\dag b \nonumber
\\
&+& \frac{\left|\Omega_{r1}\right|^2}{4\Delta_r} \left( \frac{N_1}{2}+J_{z1} \right)
+  \frac{\left|\Omega_{s1}\right|^2}{4\Delta_s} \left( \frac{N_1}{2}-J_{z1} \right)
+ \frac{\left|\Omega_{r2}\right|^2}{4\Delta_r} \left( \frac{N_2}{2}+J_{z2} \right)
+  \frac{\left|\Omega_{s2}\right|^2}{4\Delta_s} \left( \frac{N_2}{2}-J_{z2} \right)
\nonumber
\\
&+& \left[ a^\dag \left( \beta_{r1}J_1^- + \beta_{r2}J_2^+\right) + \textrm{H.c.}
\right]
+ \left[ b^\dag \left( \beta_{s1}J_1^+ + \beta_{s2}J_2^-\right) + \textrm{H.c.}
\right] .
\end{eqnarray}
\end{widetext}
Here, $\delta_{a,b}=\omega_{a,b}-(\omega_{Ls}-\omega_1)$ are detunings
of the cavity modes from Raman resonance, and
\begin{eqnarray}
\beta_{ri} = \frac{\Omega_{ri}g_{ai}^\ast}{2\Delta_r} , ~~~
\beta_{si} = \frac{\Omega_{si}g_{bi}^\ast}{2\Delta_s} ~~~ (i=1,2)
\end{eqnarray}
are the Raman transition rates. 

In the Holstein-Primakoff representation \cite{Holstein40}, 
the collective atomic operators may
be associated with harmonic oscillator annihilation and creation operators 
$c_i$ and $c_i^\dag$ ($[c_i,c_i^\dag]=1$) via
$J_i^-=(N_i-c_i^\dag c_i)^{1/2}c_i$ and $J_{zi}=c_i^\dag c_i-N_i/2$.
For the states that we aim to prepare, the mean number of atoms transferred
to the state $|1\rangle$ in each ensemble is expected to be much smaller than 
the total number of atoms, i.e., $\langle c_i^\dag c_i\rangle\ll N_i$.  
The collective atomic operators are thus well-approximated by
$J_i^-\simeq N_i^{1/2}c_i$ and $J_{zi}\simeq -N_i/2$, 
and we can reduce $H_{\rm eff}$ to the form
\begin{eqnarray}
H_{\rm eff} &=& \left( \delta_a + \frac{N_1|g_{a1}|^2}{\Delta_r} \right)a^\dag a
+  \left( \delta_b + \frac{N_2|g_{b2}|^2}{\Delta_s} \right)b^\dag b \nonumber
\\
&+& \left[ a^\dag \left( \sqrt{N_1}\beta_{r1}c_1 + \sqrt{N_2}\beta_{r2}c_2^\dag\right) 
+ \textrm{H.c.} \right] \nonumber
\\
&+& \left[ b^\dag \left( \sqrt{N_1}\beta_{s1}c_1^\dag + \sqrt{N_2}\beta_{s2}c_2\right) 
+ \textrm{H.c.} \right] ,
\end{eqnarray}
where we have omitted constant energy terms.
With appropriate choices of detunings and/or laser intensities, we assume that
the following conditions can be satisfied: 
(i) $\delta_a+N_1|g_{a1}|^2/\Delta_r=\delta_b+N_2|g_{b2}|^2/\Delta_s=0$, 
(ii) $\sqrt{N_1}\beta_{r1}= \sqrt{N_2}\beta_{s2}\equiv\beta$, and
(iii) $\sqrt{N_1}\beta_{s1}= \sqrt{N_2}\beta_{r2}\equiv r{\rm e}^{{\rm i}\theta}\beta$, 
with $r\in [0,1]$ real.
The effective Hamiltonian thus becomes
\begin{eqnarray} \label{eq:Heff1cav}
H_{\rm eff} &=& \left[ \beta a^\dagger \left( c_1 + r{\rm e}^{{\rm i}\theta} c_2^\dag \right) 
+ \textrm{H.c.} \right] \nonumber \\
&& + \left[ \beta b^\dagger \left( c_2 + r{\rm e}^{{\rm i}\theta} c_1^\dag \right) 
+ \textrm{H.c.} \right] .
\end{eqnarray}
Consider now a unitary transformation 
$\tilde{\rho}=S_{12}^+(\epsilon)\rho S_{12}(\epsilon)$ 
with the two-mode squeezing operator 
$S_{12}(\epsilon)=\exp(\epsilon^\ast c_1c_2-\epsilon c_1^\dag c_2^\dag )$, 
where
$\epsilon ={\rm e}^{{\rm i}\theta}\tanh^{-1}(r)$. The master equation for the 
atom-cavity system becomes
\begin{eqnarray} \label{eq:meco}
\dot{\tilde{\rho}} &=& -{\rm i}\left[ \tilde{H}_{\rm eff},\tilde{\rho}\right] 
+ \kappa_a D[a]\tilde{\rho} + \kappa_b D[b]\tilde{\rho} ,
\end{eqnarray}
where
\begin{eqnarray}
\tilde{H}_{\rm eff} = \sqrt{1-r^2}
\left[ \beta \left( a^\dagger c_1 + b^\dagger c_2 \right) + \textrm{H.c.} \right] ,
\end{eqnarray}
which simply describes a system of coupled oscillators. The steady state solution of
(\ref{eq:meco}) is the vacuum state for all oscillators. Reversing the unitary 
transformation, it follows that the steady state of the total system is a pure state, 
$\rho_{\rm ss}=|\psi\rangle\langle\psi |_{\rm ss}$, with
\begin{equation} \label{eq:psi_ss}
|\psi\rangle_{\rm ss} = \left\{ S_{12}(\epsilon)
|0\rangle_1\otimes |0\rangle_2 \right\} \otimes
|0\rangle_a\otimes |0\rangle_b ,
\end{equation}
i.e., the atomic ensembles are prepared in a two-mode squeezed state and the
cavity modes in the vacuum state.

The rate at which the state is prepared is determined by the eigenvalues 
associated with the coupled-oscillator master equation (\ref{eq:meco});
in particular, by the eigenvalue with the smallest non-zero magnitude, which is 
(taking $\kappa_a=\kappa_b=\kappa$) 
$\lambda_+=-(\kappa/2)+[(\kappa/2)^2-|\beta|^2(1-r^2)]^{1/2}$. This rate decreases
as $r\rightarrow 1$, but provided $|\beta|(1-r^2)^{1/2}\gtrsim\kappa/2$, the time
required to reach the steady state will be $\sim 2/\kappa$.

Defining ``position'' and ``momentum'' operators for the atomic modes by
$X_i=c_i+c_i^\dag$ and $P_i=-{\rm i}(c_i-c_i^\dag)$, respectively, the variances in
the sum and difference operators are, for the
state (\ref{eq:psi_ss}), given by
$V(X_1\pm X_2)=V(P_1\mp P_2)=2\exp[\mp 2\tanh^{-1}(r)]$.
Hence, entanglement between the atomic ensembles of the Einstein-Podolsky-Rosen 
(EPR) type \cite{Duan00,Julsgaard01,Einstein35} is generated. 
Given the stability of the atomic ground states, this entangled state should be
long-lived, and, using matter-light state-transfer schemes (see, e.g.,
\cite{Lukin03,Lukin00a,Phillips01,Liu01,Eisaman04}), readily
recoverable in the form of propagating light pulses in the cavity mode outputs.
That is, having prepared the atomic state and switched off all of the laser fields, the
fields $\Omega_{r1}$ and $\Omega_{s2}$ could be pulsed on in a suitable fashion 
at some later time to return all of the atoms to the state $|0\rangle$ and transfer the
states of ensembles 1 and 2 to the modes $a$ and $b$, respectively.
Alternatively, only one of $\Omega_{r1}$ and $\Omega_{s2}$ might be applied to
produce a single light pulse that would be entangled with the atomic ensemble that
has not undergone the state transfer process. This pulse could be used to establish remote
quantum communication, e.g., to teleport the state of a light field to an atomic ensemble.

This scheme is also readily simplified to produce single-mode squeezed states in a
single atomic ensemble \cite{Andre02,Sorensen02}. 
In particular, for a single cavity-confined ensemble and 
with $a$ and $b$ chosen to be the same mode, one can realize a dynamics 
described by a master equation of the form 
$\dot{\rho} = -{\rm i}\left[ H_{\rm eff},\rho\right] + \kappa_a D[a]\rho$
with
\begin{eqnarray}
H_{\rm eff} &=& \left[ \beta a^\dagger \left( c_1 + r{\rm e}^{{\rm i}\theta} c_1^\dag \right) 
+ \textrm{H.c.} \right]  ,
\end{eqnarray}
the steady state solution of which is $\{S_1(\epsilon)|0\rangle_1\}\otimes |0\rangle_a$
where $S_1(\epsilon)=\exp[\epsilon^\ast c_1^2-\epsilon (c_1^\dag )^2]$.

For a potential experimental system and set of parameters, we
consider ensembles of $N\sim 10^6$ ${}^{87}\textrm{Rb}$ atoms with the states 
$|0\rangle$ and
$|1\rangle$ corresponding to the ground magnetic states $\{ F=1,m_F=\pm 1\}$.
These are coupled via Raman transitions involving 
circularly-polarized ($\sigma^\pm$) cavity modes and laser fields in a ring cavity
configuration. An external magnetic field can be used to lift the degeneracy of
the $m_F=\pm 1$ states and enable distinct Raman channels between these states
\cite{Kuhn05}.
For the single-atom single-photon dipole coupling strength in a ring cavity we choose
$g/(2\pi)\sim 50~{\rm kHz}$ \cite{Kruse03,Nagorny03}, 
and assume laser Rabi frequencies
$\Omega/(2\pi)\sim 1~{\rm MHz}$ and atomic excited state detunings 
$\Delta/(2\pi)\sim 250~{\rm MHz}$ (for simplicity, we omit subscripts from the 
parameters). These give a Raman transition rate $\beta/(2\pi)\sim 100~{\rm kHz}$,
and for $r=0.8$ (giving $V(X_1+X_2)=0.22$, 
i.e., a $9.5~{\rm dB}$ reduction in the variance \cite{ndpatransfer}), 
one has $\beta(1-r^2)^{1/2}/(2\pi)\sim 60~{\rm kHz}$. Choosing 
$\kappa/(2\pi)\sim 120~{\rm kHz}$, the timescale for the state preparation is then
$\lambda_+^{-1}\sim 2/\kappa\sim 3~\mu{\rm s}$. 

The state preparation dynamics involves only the ``symmetric'' atomic modes represented
by $c_{1,2}$; readout of the atomic quantum memory is accomplished by 
coupling once more to these modes alone and adiabatically mapping their states onto the 
readout light fields. Under such circumstances the rate of decoherence of the atomic
quantum memory due to atomic spontaneous emission is given by the rate of
{\it single-atom} spontaneous emission \cite{Mewes05,SupplementDuan01}, which is
estimated here by $\gamma (\Omega^2/4\Delta^2)\sim 0.02~{\rm kHz}$, where 
$\gamma/(2\pi)\sim 6~{\rm MHz}$ is the excited state linewidth for 
${}^{87}\textrm{Rb}$. Hence, spontaneous emission should have a negligible effect
on the fidelity of the quantum memory.

Another issue to consider is uncertainty in the atom numbers $N_{1,2}$, which
could make it difficult to precisely satisfy the conditions (i--iii) for
zeroing detunings and fixing the relative Raman transition rates in the 
two ensembles. If conditions (ii--iii) are not satisfied then the steady state of the
system is no longer a pure state. 
Numerical simulations show, however, that the reduction in the EPR variance is 
degraded (for $r=0.8$) by only 1--2~dB for deviations of the ratio 
$\sqrt{N_2/N_1}\beta_{s2}/\beta_{r1}$ from unity by 10--15\%. 
We note also that if conditions (ii--iii) are not satisfied then the steady states of
the cavity modes are no longer the vacuum state and a finite output photon flux is 
expected. This output flux could in principle be monitored and 
laser detunings and/or intensities adjusted so as to zero the flux and
thereby achieve conditions (ii--iii) without exact initial knowledge of $N_{1,2}$.

\begin{figure}[h]
\begin{center}
\includegraphics[width=3.0in]{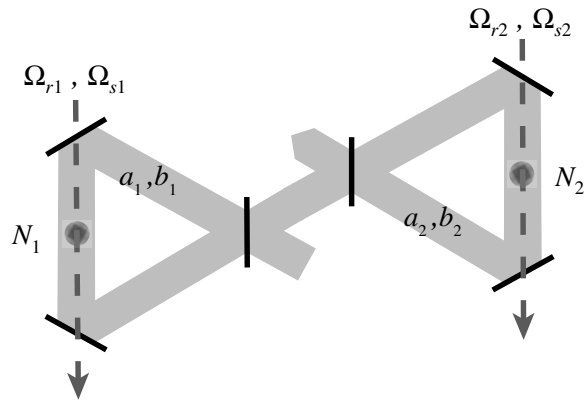}\\
\caption{
Possible cascaded ring cavity setup for the preparation of entangled 
distantly-separated atomic ensembles.
}
\label{fig:cascade}
\end{center}
\end{figure}

As mentioned earlier, matter-light state mapping schemes could be applied to 
transfer the entanglement from one of the ensembles to a propagating light pulse, 
which could in turn be used to distribute entanglement between 
distantly-separated ensembles. Alternatively, and somewhat remarkably,
the scheme described above can in fact be applied to atomic ensembles
in {\em separate}, {\em cascaded} optical cavities, as depicted in 
Fig.~\ref{fig:cascade}.

Under precisely the same conditions as were applied in deriving (\ref{eq:Heff1cav}),
the master equation for the two-cavity system takes the form
\begin{equation} \label{eq:rhodotcasc}
\dot{\rho} = -{\rm i}\left[ H_{\rm eff},\rho\right] + {\cal L}\rho ,
\end{equation}
where now
\begin{eqnarray}
H_{\rm eff} &=& \left[ \beta \left( a_1^\dag  c_1 
+ r{\rm e}^{{\rm i}\theta} a_2^\dag c_2^\dag \right)
+ \textrm{H.c.} \right] \nonumber \\
&& + \left[ \beta \left( b_2^\dag c_2 + r{\rm e}^{{\rm i}\theta} b_1^\dag c_1^\dag 
\right) + \textrm{H.c.} \right] ,
\end{eqnarray}
and the cascaded cavity dynamics are described by
\cite{Gardiner04}
\begin{eqnarray}
{\cal L}\rho &=& \kappa D[a_1]\rho + \kappa D[b_1]\rho +
 \kappa D[a_2]\rho + \kappa D[b_2]\rho \nonumber
\\
&-& 2\kappa\sqrt{\eta} \left( \left[ a_2^\dag ,a_1\rho\right] 
+ \left[ \rho a_1^\dag ,a_2\right] \right) \nonumber
\\
&-& 2\kappa\sqrt{\eta} \left( \left[ b_2^\dag ,b_1\rho\right] 
+ \left[ \rho b_1^\dag ,b_2\right] \right) .
\end{eqnarray}
Here $\eta\in [0,1]$ is the coupling efficiency between the two cavities (assumed
the same for both modes) and we have assumed the same field decay
rate $\kappa$ for all cavity modes.

Solutions to (\ref{eq:rhodotcasc}) are generally complicated
and exhibit correlations between all six modes. A simple solution arises however
in the limit $\kappa\gg|\beta|$, whereby the cavity modes can be adiabatically 
eliminated from the dynamics to leave a master equation for the 
reduced density operator $\rho_{\rm a}$ of the atomic modes alone \cite{Clark03}. 
Applying the unitary transformation 
$\tilde{\rho}_{\rm a}=S_{12}^+(-\epsilon)\rho_{\rm a} S_{12}(-\epsilon)$ and 
assuming ideal inter-cavity coupling ($\eta=1$), this
master equation reduces to the simple form
\begin{equation} \label{eq:adME}
\dot{\tilde{\rho}}_{\rm a} = \frac{|\beta|^2(1-r^2)}{\kappa} \left( D[c_1]\tilde{\rho}_{\rm a}
+ D[c_2]\tilde{\rho}_{\rm a} \right) ,
\end{equation}
so once again the steady state of the atomic system is a pure two-mode squeezed
state 
$|\psi\rangle_{\rm ss}=S_{12}(-\epsilon)|0\rangle_1\otimes |0\rangle_2$.
This steady state is produced at a rate $\Gamma=|\beta|^2(1-r^2)/\kappa$, 
which, using
parameter values as earlier [$\beta/(2\pi)\sim 100~{\rm kHz}$, $r=0.8$], but now
with $\kappa/(2\pi)\sim 500~{\rm kHz}$, takes a characteristic value
$\Gamma/(2\pi)\sim 7~{\rm kHz}$ ($\Gamma^{-1}\sim 22~\mu {\rm s}$).

In the presence of coupling loss ($\eta <1$) the steady state is mixed and the amount
of reduction in the EPR variance is limited. In particular, for
$\theta=0$ one finds
\begin{eqnarray}
V(X_1-X_2) = V(P_1+P_2) = 2\left( \frac{r^2-2r\sqrt{\eta}+1}{1-r^2} \right),
\end{eqnarray}
which takes a minimum value of $2\sqrt{1-\eta}$ for 
$r=(1-\sqrt{1-\eta})/\sqrt{\eta}$. It follows from this result that efficient coupling and
transfer between the cavities is essential for generating high degrees of 
steady state entanglement, although we note that variations on the scheme 
presented here which utilise single Raman channels and fixed-time evolution 
may enable reductions in the EPR variance below the value $2\sqrt{1-\eta}$
\cite{Peng02}.

In conclusion, we have proposed schemes for the unconditional preparation of EPR-type
entangled states of collective atomic modes in physically separated atomic ensembles.
These schemes appear within reach of current experiments and 
expand the range of possibilities for state preparation in atomic ensembles
and for remote quantum communication.

\acknowledgments
A.S.P. gratefully acknowledges support from the Alexander von Humboldt Foundation and 
from the Marsden Fund of the Royal Society of New Zealand.
ES acknowledges financial support from EU through RESQ project.


\begin{thebibliography}{13}
\bibitem{tuning}
This choice of laser frequencies allows a unitary transformation to be made that removes
all time dependence from the Hamiltonian describing the atom-cavity system.
\end{thebibliography}
\end{document}